# Sense of Community: How Important is this Quality in Blended Courses

Maryam Tayebinik and Marlia Puteh

Universiti Teknologi Malaysia

Email: ttayebi@gmail.com

**Abstract**

Combining online classes with the traditional classes foster the advantages of both learning environments. The aim of this study was to examine the effects of integrating face-to-face classes in fully online courses. Forty eight undergraduate students studying at the e-learning center of a public university in Iran were the subjects of this study. They were required to provide their feedback on the inclusion of face-to-face component in their e-learning classes. Data collected through open-ended questions indicated that the most dominant outcome of such a hybrid course on the students was the perception on the sense of community. The findings suggest that students' high satisfaction on blended learning courses was due to the fact that it promoted their sense of community. This supports another conclusion of the study that face-to-face classes and online classes are complementary and provide a balanced pedagogical role for each other.

## Introduction and Background

The introduction of Information Communication Technology (ICT) in education has resulted in a profound application of virtual instruction in many higher education institutions. Compared to the traditional classes, the predominant privilege of virtual education is its accessibility [1]. Apart from that, another noticeable dimension of virtual education is its advantage in bringing education to anyone and anywhere [2, 3]. Despite these apparent benefits, virtual education has gradually caused a sense of remoteness among distance learners. This has weakened the psychological sense of community crucial for students' satisfaction, performance and persistence [4]. Furthermore, geographical and transactional separation of students in virtual environments might develop a sense of isolation and disconnection and consequently, making it impossible to develop the sense of community in students [5].

Based on some studies, burnout [6] or dropouts [7] in online classes are highly related to the expansion of a sense of isolation and the lack of a sense of community. One of the uniqueness of traditional classes is face-to-face interaction, and this is a similar element needed in distance education in promoting the sense of community among the students. This was also the rationale of the formation of blended learning or hybrid learning [8]. In this type of learning, the potentials of both distance education and traditional classes as well as the sense of community have been incorporated. In fact, a strong sense of community in online education can boost collaboration, information exchange, learning support, satisfaction with group efforts, and responsibility to fulfil group goals [9, 10, 11, 8].

The following literature deals with the different dimension of sense of community in virtual educational settings such as student's success, satisfaction and the removal of the sense of isolation among students.

**Sense of Isolation in Distance Education**

Feeling of loneliness experienced by students when not belonging to a group or the feeling of a separate part of something can be classified as students' isolation [12].

[8] emphasized that the increased feeling of isolation might result in students' sense of alienation. Besides, there is likelihood that distance learners feel alienated and isolated because of the physical separation from other students and their distance from school and its services [13]. In this regard, [14] claimed that the feeling of isolation is considered as a source of lower satisfaction, which in turn might produce poor learning outcomes.

In another perspective, the students' perceived psychological and communication space between instructors and learners is known as transactional distance [15]. Advocates of community development in online environments rationalized that the reduction of transactional distance has promoted the expansion of online community in education [16].

According to [17] the sense of isolation in online courses has greatly affected the students and caused them to experience stress during the course. "Students reported confusion, anxiety, and frustration due to the perceived lack of prompt or clear feedback from the instructor, and from ambiguous instructions on the course website and in e-mail messages from the instructor" [17, pp. 68].

The above analysis has demonstrated that a sense of isolation in distance education is evident and its effect on the teaching and learning process is embryonic. Minimizing its consequence via maximizing the sense of community in students therefore, seems necessary to assist learning.

**What is Sense of Community?**

[18]'s definition of psychological sense of community is the reference point for the majority of researches on sense of community. The authors defined the sense of community following this simple definition: "Sense of Community is a feeling that members have of belonging, a feeling that members matter to one another and to the group, and a shared faith that members' needs will be met through their commitment to be together" [18, p. 9]. They proposed that sense of community comprises of four elements below:

- Membership: a sense of belonging and identification in a sense of security.
- Influence: group members influence each other to participate in group activities.
- Integration and fulfillment of needs: being together with the members and perceived similarity to others and this homogeneity is effective to group cohesion and interaction.
- Shared emotional connection: as a result of participation in the community.

In technology-mediated classrooms, the instructor and students interact in formal and informal settings and may not be located in the same venue. However, in traditional classes learners and instructors generally interact by confronting each other in similar location [19]. Having said this, it can be implied that distance education programs lacks sense of community because it does not promote togetherness among students. That is why some researchers believe that online students need a high sense of community in order to be academically successful whereas traditional students do not require it because of the inherent qualities of face-to-face classrooms [9].

**Blended Learning and Sense of Community**

Previous studies on distance education through computer technology [20, 21, 22] described distance education as promoting learning while teacher and learner are physically separated.

The integration of human interaction to online learning has demonstrated that the educators have considered the need for socialization, which will assist the process of learning [23]. Blended learning was the result of such a challenge through which according to [24], the potentials of web-based training with those of face-to-face classroom are combined and this is referred as hybrid learning. Likewise, [25] believed that blended learning environments possess the values of traditional classes which improve the effectiveness of meaningful learning experiences. On the other hand, the implementation of blended courses has encouraged a positive effect on the reduction of the number of exams dropouts. This has justified that blended learning activities has a positive effect on students' final grades [26].

Blended learning environments have seized the potentials of traditional classrooms through the adaptation of the online learning community. It is expected that in such education setting, the sense of community would evolve more than pure online learning environment. In order to verify such a premise, [27] conducted a comparative study to investigate the relationship of the sense of community between fully online learning, blended learning and traditional classroom in higher education learning environments. The findings suggested that the sense of community among blended learning students is stronger than students from the traditional or fully online environments.

This literature review led us to the conclusion that the sense of community is more evident in blended learning classes by comparing it with other educational settings. However, the studies was lacking investigate on the context that the same subjects experience both online learning and blended learning. The researchers, therefore, arranged a qualitative research to evaluate the significance of sense of community on a context in which students were studying in a fully online environment before amendment of the university plan to mix face-to-face classes.

## Method

In this research, a qualitative approach to study the effect of integrating some traditional elements to fully online courses was conducted. Students from the e-learning centre of the Khaje Nasir Toosi University (KNTU) of Technology in Iran were the subject of this study.

### *Procedure*

Some face-to-face classes were integrated to an online instruction in order to boost the pedagogical process. It is also carried out to fulfill the e-learners' difficulties in authentic interactive environments. Following the adjustment of the e-learning program, the effectiveness of the plan and the students' views and satisfactions were examined.

### *Subjects*

Participants of this study were 48 undergraduate students majoring in Industrial and Computer engineering. They had been using the e-learning system in almost all courses including English for Specific Purpose (the course that was thought by the researcher), in the e-learning centre of KNTU.

### *Data Collection Process*

Data was collected through open-ended questions. The students were asked to express their opinions towards the inclusion of face-to-face classes into their e-learning program. These questions were conducted three times in three different classes of ESP. These open-ended questions were:

- What do you think about the new environment of the course?
- State your opinions on the implementation of face-to-face classes along side with the online classes.
- Explain briefly about your interactions in this hybrid environment.

## Findings and Discussion

The answers to the questions were analyzed, and then the students' opinions were classified as follows:

• Genuine communication in the hybrid classes is available with the emphasis on the sense of belonging to a community. This has removed our sense of privacy and isolation.
• We could have more chances to solve our problems in face-to-face class.
• We could ask more questions from each other and make comments when engaged in this new environment.
• Student-instructor interaction is highly satisfying.
• Implementation of face-to-face classes is helpful to overcome the weaknesses of e-learning environment such as connection difficulties, quality of voice and the speed of the Internet speed.

Table 1 presents the most frequent answers to the above-mentioned questions and their percentages.

| Type of answer | Percentage |
|---|---|
| High perception of the sense of community | 42% |
| More effective student- instructor interaction | 23% |
| Positive view of the hybrid courses | 19% |
| Increase of student-student interaction | 11% |
| Others | 5% |

Table 1: *Students' Feedback and Percentages*

According to Table 1, the majority of the respondents (42%) declared that they felt belonged to a community when the hybrid courses were implemented. The introduction of the blended learning environment has also created a more effective interaction between the students and their instructor. 23 percent of the students agreed to this and they attributed the instructor's quick feedback as significant for them. They agreed that the valuable communication with the instructors have enabled them to solve mathematical problems or to better grasp the formula. 19 percent of the students were positive to the introduction of the hybrid classes and showed pleasure with the new environment. The students (11%) also viewed that their interactions with their fellow classmates have also improved through the blended learning environment.

**Conclusion**

The notions of the sense of community and blended learning environments have reciprocal relations. In other words, high satisfaction of blended learning courses is attributed to its promotion of sense of community. This supports the claim that face-to-face component and online classes are complementary and provide a balanced role for each other. However, the success of blended learning is not only due to the integration of ICT with the face-to-face approach [28] but evidence suggests that blended courses create a stronger sense of community among the students. Thus, the sense of isolation as asserted by some researchers may be the reason that some online courses recorded more university dropouts.

This has necessitated the need for curriculum reviews and adjustment on the instructional practices in order to improve student learning and to promote student growth. If the students feel that they belong to a valuable and worthwhile activity when connected to each other in the online learning environment, there is a high possibility that this could reduce their feelings of alienation hence, increase their sense of community within the context of a virtual classroom.